\def\a{\alpha}\def\b{\beta}\def\d{\delta}\def\e{\epsilon}
\def\f{\phi}\def\h{\theta}
\def\l{\lambda}\def\m{\mu}\def\n{\nu}\def\o{\omega}\def
\p{\pi}\def\q{\psi}
\def\y{\eta}

\def\D{\Delta}

\def\de{\partial}
\def\inf{\infty}\def\mo{{-1}}\def\ha{{1\over 2}}

\def\({\left(}\def\){\right)}\def\[{\left[}\def\]{\right]}

\def\const{{\rm const}}\def\ex{{\rm e}}
\def\arcth{{\rm arctanh\,}}\def\arccth{{\rm arccoth\,}}
\def\arcsh{{\rm arcsinh\,}}\def\arcch{{\rm arccosh\,}}

\def\mn{{\mu\nu}}

\def\exp#1{{\rm e}^{#1}}

\def\fe{field equations }
\def\tran{transformations }\def\coo{coordinates }
\def\gs{ground state }

\def\cc{coupling constant }

\def\rep{representation }

\def\pb{Poisson brackets }

\def\poi{Poincar\'e }
\def\des{de Sitter }

\def \schr{Schr\"odinger }

\def\eom{equations of motion }
\def\cor{commutation relations }
\def\ev{expectation value }

\def\section#1{\bigskip\noindent{\bf#1}\smallskip}

\def\nota{\footnote{$^\dagger$}}
\font\small = cmr8

\def\PL#1{Phys.\ Lett.\ {\bf#1}}

\def\PR#1{Phys.\ Rev.\ {\bf#1}}

\def\JoP#1{J.\ Phys.\ {\bf#1}} \def\IJMP#1{Int.\ J. Mod.\ Phys.\ {\bf #1}}
 
\def\PRep#1{Phys.\ Rep.\ {\bf#1}}

\def\JHEP#1{JHEP\ {\bf#1}}
\def\RMP#1{Rev.\ Mod.\ Phys.\ {\bf#1}}\def\AdP#1{Annalen Phys.\ {\bf#1}}
\def\hep#1{{\tt hep-th/#1}}

\def\ref#1{\medskip\everypar={\hangindent 2\parindent}#1}
\def\beginref{\begingroup
\bigskip
\centerline{\bf References}
\nobreak\noindent}
\def\endref{\par\endgroup}

\input epsf
\def\bp{{\bf p}}
\def\pref{\sqrt{1+\b^2P^2}}\def\pres{\sqrt{1-\b^2P^2}}\def\pren{\sqrt{\b^2P^2-1}}
\def\prep{\sqrt{1-\b^2P_r^2}}
\def\dep{{\de\over\de P}}\def\bra{\langle}\def\ket{\rangle}\def\brah{\langle\hat }
\def\hx{\hat x}\def\hL{\hat L}\def\bp{\bar p}
\def\psc{phase space coordinates }\def\ur{uncertainty relations }
\font\smallmath = cmmi7\font\small = cmr7

\magnification=1200

{\nopagenumbers
\line{}
\vskip40pt
\centerline{\bf Classical and quantum mechanics of the nonrelativistic Snyder model}
\vskip40pt
\centerline{{\bf S. Mignemi}\nota{e-mail:smignemi@unica.it}}
\vskip10pt
\centerline {Dipartimento di Matematica, Universit\`a di Cagliari}
\centerline{viale Merello 92, 09123 Cagliari, Italy}
\smallskip
\centerline{and INFN, Sezione di Cagliari}
\vskip60pt
\centerline{\bf Abstract}
\bigskip
{\noindent 
The Snyder model is an example of noncommutative spacetime admitting a fundamental 
length scale $\b$ and invariant under Lorentz transformations, that can be 
interpreted as a realization of the doubly special relativity axioms.
Here, we consider its nonrelativistic counterpart, i.e.\ the Snyder model restricted 
to three-dimensional Euclidean space. 
We discuss the classical and the quantum mechanics of a free particle in this framework, 
and show that they strongly depend on the sign of a \cc $\l$, appearing in the 
fundamental commutators and proportional to $\b^2$. 
For example, if $\l$ is negative, momenta are bounded. On the contrary, for positive 
$\l$, positions and areas are quantized.
We also give the exact solution of the harmonic oscillator equations both in the
classical and the quantum case, and show that its frequency is energy dependent.
}
\vskip120pt
P.A.C.S. Numbers: 02.40.Gh; 45.20.Jj; 03.65.Ca.
\vfil\eject}

\section{1. INTRODUCTION}
Several years ago, in the attempt to introduce a short distance cutoff in field theory,
Snyder proposed a model of noncommutative spacetime, admitting a fundamental length scale 
and invariant under the Lorentz group [1].
This proposal was ahead of its time and went almost unnoticed, until a few years ago, 
when noncommutative spacetimes became fashionable, mainly in connection with string
theories, where they emerge in certain low-energy limits [2]. Also considerations on 
quantum gravity and black hole physics seem to indicate that the structure of spacetime 
must be noncommutative at scales close to the Planck length [3]. In particular, the
extremely high energies necessary to resolve very small distances could perturb the 
spacetime structure by their quantum gravitational effects.

Similar arguments on the structure of spacetime at small length scales were also the basis 
of the proposal of doubly special relativity (DSR) [4]. 
This is a model of spacetime admitting a fundamental scale that sets a bound 
on the allowed values of the momentum, and implies a deformation of the \poi symmetry and
of the dispersion relations of elementary particles. 
The natural realization of DSR is on a phase space equipped with a noncanonical symplectic 
structure [5]. It turns out that the Snyder model can be interpreted as an instance of DSR. 
In [6], it was shown in fact that the Snyder algebra is a particular realization of the 
general DSR algebra. In [7], the dynamics of the Snyder model was investigated in the 
context of DSR, showing that it implies the existence of a maximal allowed value for the 
mass of a free particle.

The model of spacetime proposed by Snyder is based on the commutation relations\footnote{$^1$}
{We use the following conventions: Greek indices run from 0 to 3, Latin indices from 1 to 3, 
the metric signature is $(-,+,+,+)$.}

$$[x_\m,x_\n]=i\,\l\,J_\mn,\qquad[p_\m,p_\n]=0,\qquad[x_\m,p_\n]=
i\left(\y_\mn+\l\,p_\m p_\n\right),\eqno(1.1)$$
where $\l$ is a coupling constant, usually assumed to be of the scale of the square of the
Planck length, and the $J_\mn$ are the generators of the Lorentz algebra. In contrast with the 
most common models [3], the commutators are not constant, but are functions of the phase 
space variables, and this allows them to be compatible with the Lorentz symmetry.
The algebra (1.1) can be obtained by constraining the momenta to lie on a hypersphere in a
(4+1)-dimensional space [1]. From this point of view the Snyder model can be viewed as the 
equivalent of \des spacetime for momentum space.

In spite of its recent revival, the physical content of the Snyder model has not been 
investigated in detail. Most investigations have been in fact directed to its formal properties.
In [8] its dynamics was derived from a constrained Hamiltonian system. Using similar methods,
the authors of [9] obtained the Snyder \cor from a six-dimensional setting. In [10] the same
techniques were used to study the symmetries of the model.

The most interesting physical implications of the Snyder model are a generalization of 
the \ur [6,11], similar to that proposed in [12], implying a lower bound for the uncertainty in 
position, and the discreteness of the spectra of area and volume [13]. The interpretation of
the model in operational terms has also been addressed in [14].

Till now, all the investigations have been restricted to the case $\l>0$. However, the 
physics strongly depends on the sign of the coupling constant. In particular, the case 
$\l<0$, that we call anti-Snyder in analogy with anti-de Sitter, is the one relevant for 
DSR, since it implies an upper bound on the mass of free particles [7]. It is therefore 
interesting to compare the two possibilities.

In this paper we shall consider the properties of a 3-dimensional Euclidean version of the 
Snyder model.
This can be interpreted as a restriction of the original model to its spatial sections, or
more properly as a nonrelativistic version of the model.
The interest of limiting our considerations to the nonrelativistic model relies in the fact 
that, while the main features (noncommutativity of the geometry, generalized uncertainty 
relations) of the relativistic model are maintained, 
one can easily implement quantum mechanics, whereas the definition of a relativistic quantum
mechanics would pose nontrivial conceptual problems [14]. We plan to investigate this topic
in a future paper.

As in the relativistic case, also the properties of the nonrelativistic Snyder model strongly 
depend on the sign of the \cc $\l$. 
If $\l>0$, the momenta are allowed to take any real value, but in the 
quantum theory a minimal uncertainty in the positions arises. In particular, in the case 
of a single spatial dimension, the model reduces to the one introduced in [12] in a different 
context.
If $\l<0$, instead, the modulus of the momentum has an upper bound $1/|\l|$, but no minimal 
uncertainty occurs in the quantum theory. However, in contrast with standard quantum mechanics,
states with vanishing position uncertainty have finite momentum uncertainty.
Moreover, length, area and volume are quantized for positive $\l$, but not for $\l<0$.

Another interesting application is the study of the harmonic oscillator. Both in classical and
quantum mechanics its solution contains corrections of order $\l E$ to the standard case.
In particular, the frequency of oscillation is no longer independent from the energy.

It may also be interesting to notice the existence of an alternative realization of the Snyder
\cor for negative $\l$, that yields a lower bound for the momentum. Although this possibility 
is pathological under some respects, we briefly discuss it.

\section{2. CLASSICAL MECHANICS OF THE SNYDER MODEL}
We first consider the classical implementation of the Snyder model on phase space. In the
relativistic case this has been investigated in several papers from various points of view
[8,10,7,15].
\section{2.1 The model}
Classically, the nonrelativistic Snyder model can be realized by postulating a noncanonical 
symplectic structure, with fundamental \pb
$$\{x_i,x_j\}=\l\,J_{ij},\qquad\{p_i,p_j\}=0,\qquad\{x_i,p_j\}=\d_{ij}+\l\,p_ip_j,\eqno(2.1)$$
where $J_{ij}=x_ip_j-x_jp_i$ are the generators of the group of rotations, and the sign of the 
coupling constant $\l$ determines the properties of the model.

In analogy with its relativistic counterpart [1], the model can be derived from a 4-dimensional 
momentum space, constraining the momenta to live on a 3-dimensional hypersurface. This 
construction is the same as that of de Sitter space, but in momentum space. 
For example, the momentum space of the original Snyder model, with $\l=\b^2>0$, can be 
represented as a 3-sphere of radius $1/\b$ embedded in 4-dimensional Euclidean space of 
coordinates $P_a$, with $a=1,\dots,4$. The points on the sphere satisfy the equation 
$P_a^2=1/\b^2$. 
The construction can be extended to the full phase space [15], by introducing the 
4-dimensional coordinates $X_a$ satisfying canonical \pb with the momenta $P_a$.

Choosing projective coordinates $p_i$ on the 3-sphere,
$$p_i={P_i\over\b P_4}={P_i\over\sqrt{1-\b^2P_k^2}},\eqno(2.2)$$
where $P_k^2<1/\b^2$, with inverse transformations
$$P_i={p_i\over\sqrt{1+\b^2p_k^2}},\qquad\qquad\b P_4={1\over\sqrt{1+\b^2p_k^2}},\eqno(2.3)$$
and defining 3-dimensional position coordinates $x_i=\sqrt{1-\b^2P_k^2}\ X_i$, that 
transform covariantly with respect to the $p_i$, one obtains the \pb of the Snyder model, 
namely
$$\{x_i,x_j\}=\b^2\,J_{ij},\qquad\{p_i,p_j\}=0,\qquad\{x_i,p_j\}=\d_{ij}+\b^2\,p_ip_j.
\eqno(2.4)$$
The momentum components $p_i$ so defined range over all real values. Of course, the \tran
relating the coordinates $X_i$, $P_i$ with $x_i$, $p_i$ are not canonical.

\smallskip
Similarly, the anti-Snyder model, with $\l=-\b^2<0$, can be obtained by embedding a 
3-dimensional two-sheeted hyperboloid of equation $P_k^2-P_4^2=-1/\b^2$ in 4-dimensional 
momentum space with Minkowskian signature.
 
Choosing again projective coordinates 
$$p_i={P_i\over\b P_4}={P_i\over\sqrt{1+\b^2 P_k^2}},\eqno(2.5)$$
with inverse
$$P_i={p_i\over\sqrt{1-\b^2p_k^2}},\qquad\qquad\b P_4={1\over\sqrt{1-\b^2p_k^2}},\eqno(2.6)$$
and defining $x_i=\sqrt{1+\b^2P_k^2}\ X_i$, one obtains the \pb
$$\{x_i,x_j\}=-\b^2\,J_{ij},\qquad\{p_i,p_j\}=0,\qquad\{x_i,p_j\}=\d_{ij}-\b^2\,p_ip_j.
\eqno(2.7)$$
In this case, the momenta are bounded by the relation $p_i^2<1/\b^2$, like in 
some models of doubly special relativity.

\smallskip
In principle, the \pb (2.7) may also be derived from a one-sheeted hyperboloid 
$P_k^2-P_4^2=1/\b^2$ embedded in in 4-dimensional momentum space with Minkowskian signature.
We shall call the resulting model pro-Snyder.
In this case, the momentum space has not constant curvature, but this fact has no relevance
for our considerations. However, as we shall see, this model suffers some pathologies 
caused by the fact that in the limit $\b\to0$, the momentum becomes imaginary, so that the
kinetic energy has the wrong sign. Therefore we shall not study it in detail, but only give 
some hints on its properties.

The relations between $p_i$ and $P_i$ are in this case
$$p_i={P_i\over\sqrt{\b^2P_k^2-1}},\eqno(2.8)$$
with inverse
$$P_i={p_i\over\sqrt{\b^2p_k^2-1}},\qquad\qquad\b P_4={1\over\sqrt{\b^2p_k^2-1}},\eqno(2.9)$$
and the position \coo are defined as $x_i=\sqrt{\b^2P_k^2-1}\ X_i$.
Now, both $p_k^2$ and $P_k^2$ possess a lower bound, given by $1/\b^2$. In a relativistic
extension of the model, this would lead to a new kind of DSR, with the momenta displaying a 
lower bound. 

Summarizing, we have shown the possibility of defining three different models 
that obey the Snyder algebra, and differ in the range of 
definition of the momentum. While in the first case the momenta can take any real value,
in the other cases they have an upper or a lower bound, respectively. The latter models can
be interpreted in the framework of DSR, where analogous bounds on the allowed values of the 
momentum occur.

\section{2.2 Symmetries}
While the momentum space is by construction invariant under $SO(4)$ or $SO(3,1)$, 
depending on the sign of $\l$, from a physical point of view the spatial symmetries 
of the model, that act on the position coordinates, are more relevant. Let us therefore
consider in detail the transformation rules of the full phase space variables.

The \psc transform as vectors under the action of the generators of 
rotations $J_{ij}=x_ip_j-x_jp_i=X_iP_j-X_jP_i$,
$$\{J_{ij},x_k\}=\d_{ik}x_j-\d_{ij}x_k,\qquad 
\{J_{ij},p_k\}=\d_{ik}p_j-\d_{ij}p_k,\eqno(2.10)$$
while the \pb (2.1) transform covariantly. The symmetry under rotations is therefore 
realized in the usual way.

The \pb are instead not covariant under ordinary translations. In order to preserve 
their covariance, the translation symmetry must be realized in a nonlinear and
momentum-dependent way, as in DSR [10]. This fact gives rise to some ambiguity in the 
definition of the generators of translations [7].
The simplest choice is to identify the generators $T_i$ with the momenta $p_i$, 
leading, for a translation of infinitesimal parameter $a_i$, to the relation
$$\d x_i=a_j\,\{x_i,p_j\}=a_i+\l\,a_jp_jp_i.\eqno(2.11)$$

A different choice is possible and is particularly useful in the context of DSR, since
it gives rise to deformed dispersion relations.
In this case, one identifies the translation generators with the variables $P_i$, and
hence
$$\d x_i=a_j\,\{x_i,P_j\}={a_i\over\sqrt{1+\l p_i^2}}\,.\eqno(2.12)$$

For both choices, the momenta $p_i$ are of course unaffected by the transformation.
\bigbreak

\section{2.3 Classical motion}

The Hamiltonian for a free particle can be defined as the square of the translation 
generator. For the choice $T_i=p_i$, it has the usual form
$$H={p_i^2\over2m}.\eqno(2.13)$$
If $T_i=P_i$, instead,
$$H={P_i^2\over2m}={1\over2m}\ {p_i^2\over1+\l p_k^2}.\eqno(2.14)$$
Both expressions are invariant under rotations and translations.

The \fe are obtained taking care of the deformed Poisson structure. The Hamiltonian (2.13)
yields
$$\dot x_i=(1+\l p_k^2)\,p_i,\qquad\qquad\dot p_i=0,\eqno(2.15)$$
while (2.14) gives
$$\dot x_i={p_i\over1+\l p_k^2},\qquad\qquad\dot p_i=0.\eqno(2.16)$$
In both cases, the relation between velocity and momentum is no longer linear. However,
the solutions are the classical ones, $p_i=\const$, $x_i=\a t+\b$.

Since the second possibility is more interesting in a relativistic context, in the following 
we shall limit our considerations to the form (2.13) of the kinetic term of the Hamiltonian.
\medskip
A nontrivial example of dynamics is that of a one-dimensional harmonic oscillator, 
with Hamiltonian
$$H={p^2\over2m}+{m\o_0^2x^2\over2}.\eqno(2.17)$$
For unit mass, the Hamilton equations read 
$$\dot x=(1+\l p^2)\,p,\qquad\qquad\dot p=-\o_0^2(1+\l p^2)\,x.\eqno(2.18)$$

Let us first consider the Snyder case, $\l=\b^2>0$. The equations (2.18) can be solved as 
follows: the second one can be written
$${d(\arctan\b p)\over dt}=-\o_0^2\b x.\eqno(2.19)$$
Defining $\bar p=\arctan\b p$, deriving (2.19), and substituting the first Hamilton equation, 
one obtains
$${d^2\bp\over dt^2}=-\o_0^2{\sin\bp\over\cos^3\bp}\,,\eqno(2.20)$$
which admits the first integral
$$\ha\left({d\bp\over dt}\right)^2+{\o_0^2\over2}\ \tan^2\bp=\const=\o_0^2\b^2E,
\eqno(2.21)$$
where the integration constant $E$ has be chosen so that it coincides with the total 
energy of the oscillator. Clearly, these equations are identical to those of a classical 
particle moving in the effective potential $V=\o_0^2\tan^2\bp$, depicted in fig.\ 1.
\bigskip
\centerline{\epsfysize=5truecm\epsfbox{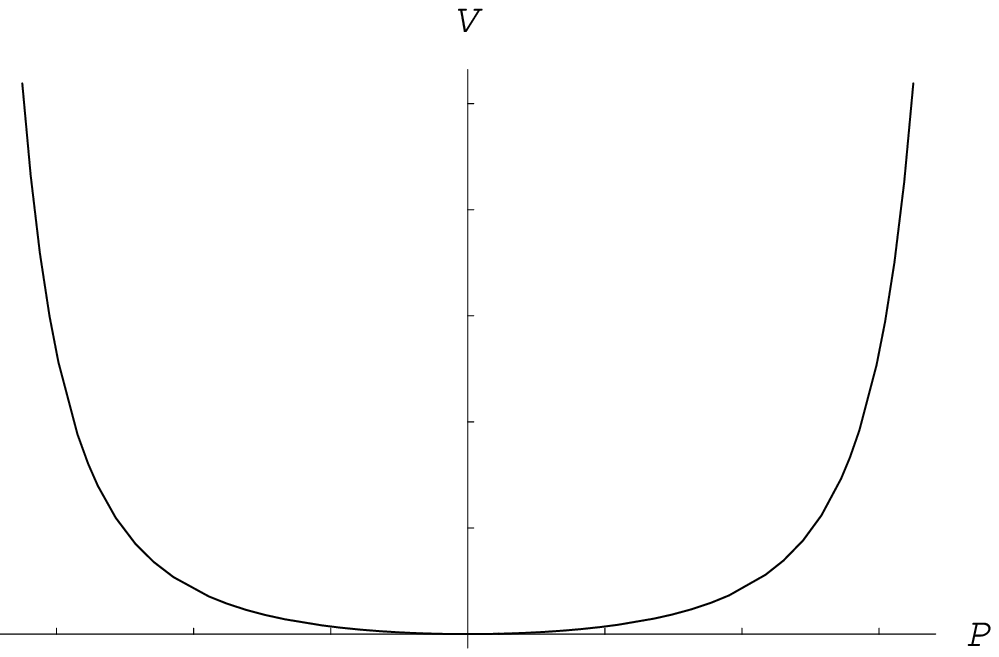}}

\centerline{\noindent\small Fig. 1: The effective potential for the Snyder oscillator.}
\bigskip
The integration of (2.21) yields
$$\sin\left(\sqrt{1+2\b^2E}\ \o_0t\right)=\sqrt{1+2\b^2E\over2\b^2E}\ \sin\bar p,\eqno(2.22)$$
and hence
$$p={\sqrt{2E}\ \sin\left(\sqrt{1+2\b^2E}\ \o_0t\right)\over\sqrt{1+2\b^2E\cos^2
\left(\sqrt{1+2\b^2E}\ \o_0t\right)}}.\eqno(2.23)$$
From (2.18) it is then easy to obtain for $x$
$$x={{\sqrt{2E(1+2\b^2E)}\ \cos\left(\sqrt{1+2\b^2E}\ \o_0t\right)\over\o_0\
\sqrt{1+2\b^2E\cos^2\left(\sqrt{1+2\b^2E}\ \o_0t\right)}}}.\eqno(2.24)$$
\bigskip
\centerline{\epsfysize=5truecm\epsfbox{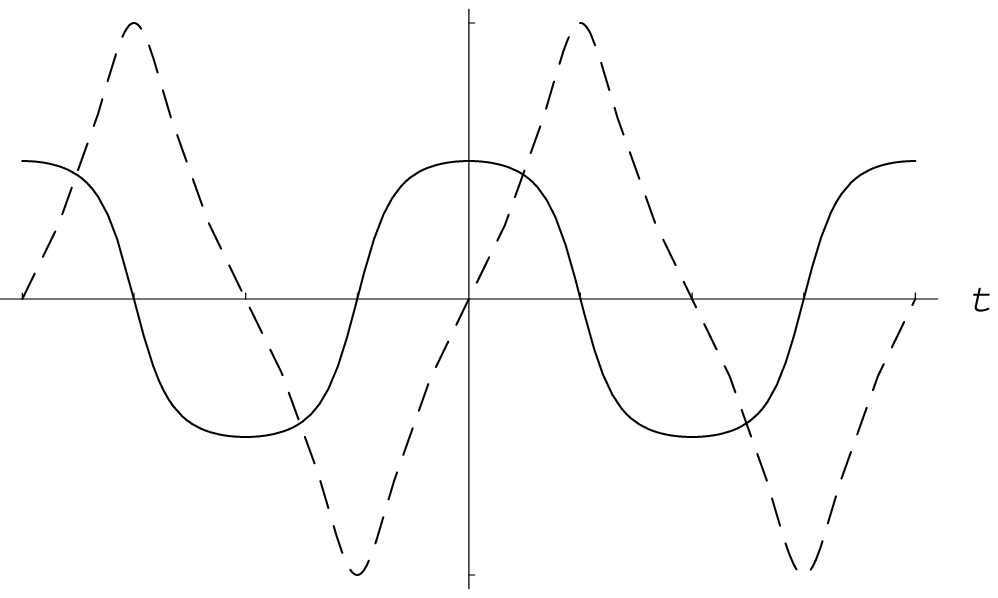}}
\medskip
\centerline{\noindent\small Fig. 2: The solution of the Snyder oscillator for 
$\scriptstyle \b^2E=3$. The solid line}
\centerline{\quad\quad\ {\small represents the coordinate} ${\scriptstyle x}${\small, 
the dashed line the coordinate} {\smallmath p}{\small.}}
\bigbreak

The harmonic oscillator has therefore a different solution than in classical mechanics. 
The solution is still periodic, but the frequency $\o$ presents energy-dependent corrections 
of order $\b^2E$, $\o=\sqrt{1+2\b^2E}\,\o_0$. Also the amplitude acquires corrections of the 
same order of magnitude and is no longer sinusoidal (see fig.\ 2). 

\medskip
In the anti-Snyder case, $\l=-\b^2<0$, the solution can be obtained in an analogous way. 
Writing the second Hamilton equation as
$${d(\arcth\b p)\over dt}=-\o_0^2\b x,\eqno(2.25)$$
and defining the variable $\bp=\arcth\b p$, one goes through the same steps as before.
In particular, one has
$${d^2\bp\over dt^2}=-\o_0^2{\sinh\bp\over\cosh^3\bp}\,,\eqno(2.26)$$
with first integral
$$\ha\left({d\bp\over dt}\right)^2+{\o_0^2\over2}\tanh^2\bp=\const=\o_0^2\b^2E,
\eqno(2.27)$$
where the integration constant $E$ has be chosen so that it vanishes at the minimum of 
the potential and coincides with the total energy of the oscillator. From the condition
$p^2<1/\b^2$, it follows that $\b^2E<\ha$. In this case the \eom coincide with 
those of a particle in the potential $V=\o_0^2\tanh^2\bp$, depicted in fig.\ 3.
\bigskip
\centerline{\epsfysize=5truecm\epsfbox{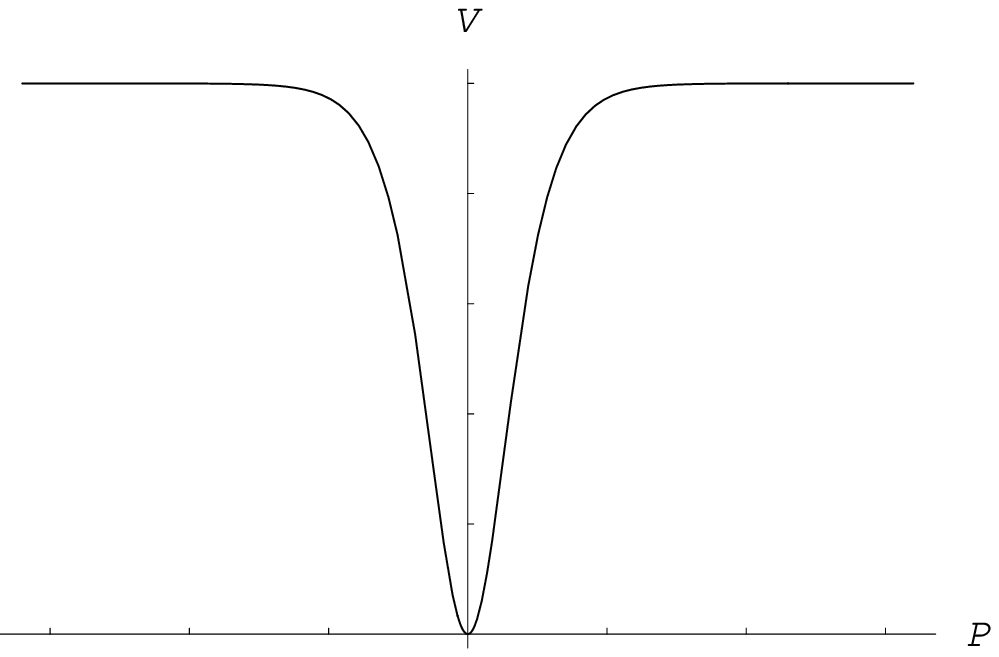}}

\centerline{\noindent\small Fig. 3: The effective potential for the anti-Snyder oscillator.}
\bigskip

After performing the integration of eq.\ (2.27), the final result turns out to be the 
analytic continuation of the solution (2.23)-(2-24) for $\b^2\to-\b^2$, namely
$$p={\sqrt{2E}\ \sin\left(\sqrt{1-2\b^2E}\ \o_0t\right)\over\sqrt{1-2\b^2E\cos^2
\left(\sqrt{1-2\b^2E}\ \o_0t\right)}},\eqno(2.28)$$
and
$$x={{\sqrt{2E(1-2\b^2E)}\ \cos\left(\sqrt{1-2\b^2E}\ \o_0t\right)\over\o\
\sqrt{1-2\b^2E\cos^2\left(\sqrt{1-2\b^2E}\ \o_0t\right)}}}.\eqno(2.29)$$
The properties of the solutions are analogous to those found in the Snyder case (see fig.\ 4).
\bigskip
\centerline{\epsfysize=5truecm\epsfbox{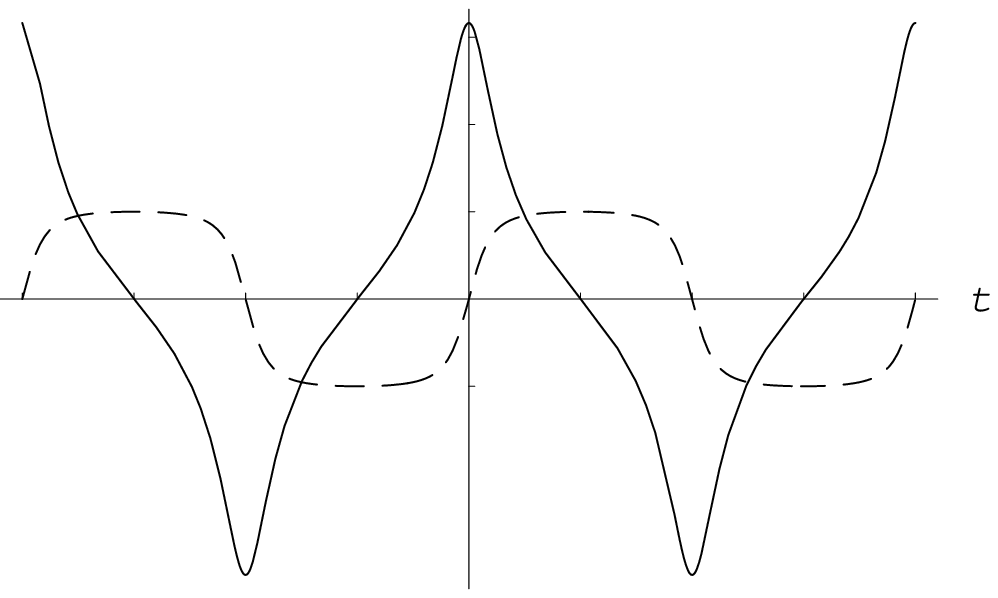}}
\medskip
\centerline{\noindent\small Fig. 4: The solution of the anti-Snyder oscillator for 
${\scriptstyle{\b^2 E}}$ = 0.9. The  solid } 
\centerline{\quad\quad\ {\small line represents the coordinate} ${\scriptstyle x}${\small, the 
dashed line the coordinate} {\smallmath p}{\small.}}

\bigskip

\smallskip
Finally, let us briefly consider the pro-Snyder case. Defining $\bp=\arccth\b p$, one 
obtains 
$$\ha\left({d\bp\over dt}\right)^2+{\o_0^2\over2}\coth^2\bp=\const=\o_0^2\b^2E,\eqno(2.30)$$
with $\b^2E>\ha$.
In this case the effective potential $V=\o_0^2\coth^2\bp$ has no minimum and 
therefore the motion is not bounded  (see fig.\ 5).
The  solution can in fact be written in terms of hyperbolic functions.
Intuitively, this behavior can be related to the fact that, as observed in sect.\ 2.1,
in the limit $\b\to0$ the kinetic energy of the pro-Snyder model has the wrong sign. 
Indeed, periodic solutions are available if $\o_0^2<0$.
\bigskip
\centerline{\epsfysize=5truecm\epsfbox{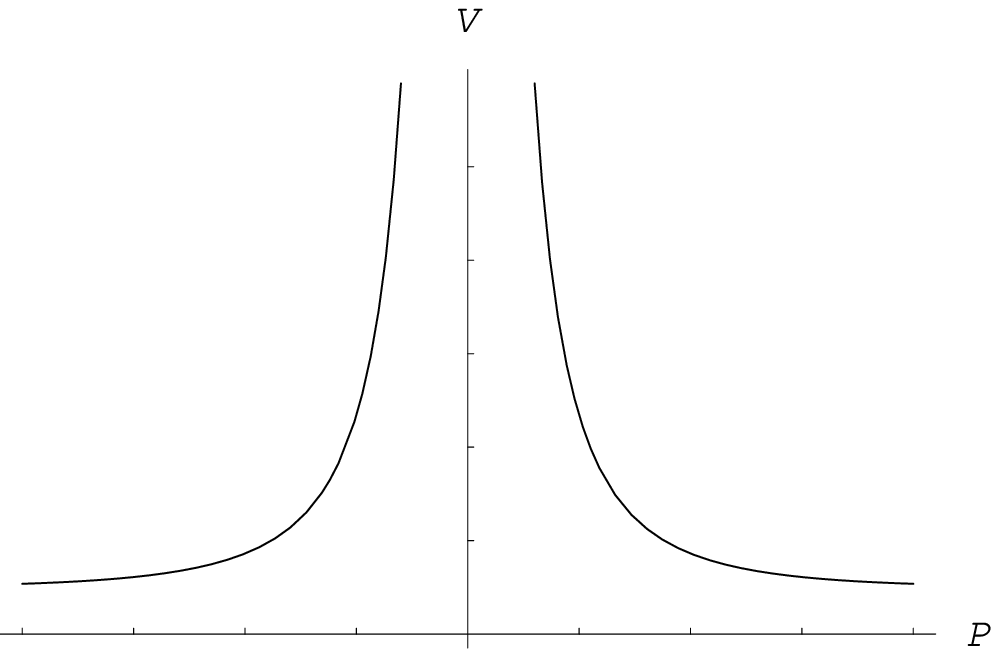}}

\centerline{\noindent\small Fig. 5: The effective potential for the pro-Snyder oscillator.}
\bigbreak

\section{3. QUANTUM MECHANICS OF THE SNYDER MODEL}
As it is well known, when passing from classical to quantum mechanics, the \pb go to commutators,
and the deformation of the latter implies a modification of Heisenberg uncertainty relations. 
Modified uncertainty relations have been studied in several papers [16]. In the following, we 
shall employ the methods introduced in [12], in order to study the specific deformation 
induced by the nonrelativistic Snyder model.
\bigbreak
\section{3.1 The Snyder model}
For Snyder space, the \cor between the position operators $\hat x_i$ and the momentum operators
$\hat p_i$ read
$$[\hat x_i,\hat x_j]=i\hbar\b^2\hat J_{ij},\qquad[\hat p_i,\hat p_j]=0,\qquad
[\hat x_i,\hat p_j]=i\hbar(\d_{ij}+\b^2\hat p_i\hat p_j).\eqno(3.1)$$
From the results of sect.\ 2, it is easy to see that they can be realized introducing 
auxiliary operators $\hat X_i$ and $\hat P_i$ obeying canonical commutation relations, and 
performing the nonlinear and nonunitary transformations
$$\hat x_i=\sqrt{1-\b^2\hat P_k^2}\ X_i,\qquad\hat p_i={\hat P_i\over\sqrt{1-\b^2\hat P_k^2}}.
\eqno(3.2)$$
The spectrum of $\hat P_i$ must be bounded by $P_k^2<1/\b^2$.

The \ur following from (3.1) are
$$\D x_i\D p_j\ge\ha\left|\,\langle[\hat x_i,\hat p_j]\rangle\right|={\hbar\over2}
\,[\d_{ij}+\b^2\D p_i\D p_j+\b^2\brah p_i\ket\brah p_j\ket],\eqno(3.3)$$
where $\bra\ \ket$ denotes expectation values.

We first consider the one-dimensional case. 
In one dimension, the \ur reduce to
$$\D x\D p\ge\ha\left|\,\langle[\hat x,\hat p]\rangle\right|={\hbar\over2}[1+\b^2(\D p)^2
+\b^2\brah p\ket^2].\eqno(3.4)$$
These generalized \ur have been thoroughly studied in ref.\ [12], where it was shown that they
imply the existence of a minimal position uncertainty, given by
$$\D x_M=\hbar \b\sqrt{1+\b^2\brah p\ket^2}.\eqno(3.5)$$
Its minimum value is obtained when $\brah p\ket=0$, as
$$\D x_0=\hbar \b,\eqno(3.6)$$
in which case $\D p=1/\b$. 

Exploiting (3.2), we can define the position and momentum operators $\hat p$ and $\hat x$
acting on functions defined on a momentum space parametrized by $P$, as
$$\hat p\,\q(P)=p\,\q(P)={P\over\pres}\ \q(P),\qquad\hat x\,\q(P)=i\hbar\pres\ 
{\de\q(P)\over\de P}.\eqno(3.7)$$
The range of allowed values of $P$ is bounded by $P^2<1/\b^2$, but the spectrum of the momentum 
operator $\hat p$ is unbounded. In [12] a different representation of the same \cor was 
discussed.

In order to define symmetric operators, i.e.
$$(\hat p\,\q,\f)=(\q,\hat p\,\f),\qquad(\hat x\,\q,\f)=(\q,\hat x\,\f),\eqno(3.8)$$
the scalar product must be defined as
$$(\q,\f)=\int_{-1/\b}^{1/\b}{dP\over\pres}\ \q^*(P)\,\f\,(P),\eqno(3.9)$$
and the wave functions must be such that $\f(1/\b)=\f(-1/\b)$.
In fact, 
$$\eqalign{&\int_{-1/\b}^{1/\b}{dP\over\pres}\ \q^*(P)\,i\hbar\pres\ \de_p\f\,(P)\cr
&=i\hbar\,\q^*\f\ \Big|^{1/\b}_{-1/\b}+\int_{-1/\b}^{1/\b}{dP\over\pres}\ 
[i\hbar\pres\ \de_p\q\,(P)]^*\f(P).}\eqno(3.10)$$

We pass now to study the spectrum of the position operator.
The position eigenfunctions $\q_x$, with eigenvalue $x$, are determined by the equation
$$i\hbar\pres\ {\de \q_x\over\de P}=x\,\q_x,\eqno(3.11)$$
whose solution is 
$$\q_x=C\ {\rm exp}\left[-{ix\over\hbar\b}\,\arcsin\b P\right].\eqno(3.12)$$

The solutions are normalizable, since
$$|\q_x|^2=|C|^2\int_{-1/\b}^{1/\b}{dP\over\pres}={\p\over\b}|C|^2.\eqno(3.13)$$
and one can set $C=\sqrt{\b\over\p}$. Moreover, the boundary term in (3.10) cancels 
only if $x=2n\hbar\b$, with integer $n$, and the spectrum of the position operator
is therefore discrete.

As discussed in [12] for a different representation, these states are not physical, since 
the \ur (3.4) imply that there are no states with a definite value of the position. 
Indeed, the expectation values of the momentum $\hat p$ and of the kinetic energy 
$\hat p^2/2m$  in the states (3.12) diverge. One can nevertheless calculate the scalar 
product,
$$(\q_x,\q_{x'})={\b\over\p}\int_{-1/\b}^{1/\b}{{\rm exp}\left[-{i(x-x')\over\hbar\b}\,
\arcsin\b P\right]\over\pres}\ dP={\hbar\b\over\pi(x-x')}\sin\left[{\pi(x-x')\over2\hbar\b}
\right],\eqno(3.14)$$
which shows that the states satisfying the correct boundary conditions are orthogonal.

\smallskip 
As proposed in [12], a more relevant basis for the position operator can be obtained by 
considering states of maximal localization $\f_x$, i.e.\ states having minimal uncertainty 
$\D x_0$ around the position $x$.
These states satisfy the equation 
$$\left(\hat x\,-\brah x\ket+{\bra[\hat x,\hat p]\ket\over2(\D p)^2}\,(\hat p\,-\brah p\ket)
\right)\f_x=0,\eqno(3.15)$$
and are obtained when $\brah p\ket=0$, $\D p=1/\b$. Therefore, in our representation,
they obey the differential equation
$$\left(i\hbar\pres\ \dep-x+{i\hbar\b^2 P\over\pres}\right)\phi_x=0,\eqno(3.16)$$
which has solution
$$\phi_{x}=N\,\pres\ \ex^{-{ix\over\hbar\b}\,\arcsin\b P},\eqno(3.17)$$
with normalization constant $N=\sqrt{2\b\over\p}$. Also these states must satisfy the 
condition $x=2n\hbar\b$ in order to define a symmetric position operator, but
contrary to the functions (3.12), they have finite \ev of the momentum and 
the kinetic energy, and are therefore physically meaningful.

The orthogonality properties of this basis can be obtained by calculating the integral
$$(\f_x,\f_{x'})={2\b\over\p}\int_{-1/\b}^{1/\b}\pres\ \exp{-{i(x-x')\over\hbar\b}\,
\arcsin\b P}\,dP$$
$$={2\hbar\b\over\p(x-x')}\ \sin{\p(x-x')\over2\hbar\b}\ \left[1-\left({\p(x-x')\over2\hbar\b}
\right)^2\right]^\mo.\eqno(3.18)$$
Curiously, this result is the same as that obtained in [12] using a different representation 
of the operators. However, like the states (3.12), also the maximally localized states that
satisfy the correct boundary conditions are orthogonal.
\bigskip

\section{3.2 The anti-Snyder model}
When $\l=-\b^2<0$, the \cor read
$$[\hat x_i,\hat x_j]=-i\hbar\b^2\hat J_{ij},\qquad[\hat p_i,\hat p_j]=0,\qquad
[\hat x_i,\hat p_j]=i\hbar(\d_{ij}-\b^2\hat p_i\hat p_j).\eqno(3.19)$$
and, in analogy with the previous case, can be realized by defining
$$\hat x_i=\sqrt{1+\b^2\hat P_k^2}\ \hat X_i,\qquad \hat p_i={\hat P_i\over
\sqrt{1+\b^2\hat P_k^2}},\eqno(3.20)$$
where $\hat X_i$ and $\hat P_i$ obey canonical commutation relations. From the definition 
follows that the spectrum of the momentum is bounded from above, $p_k^2<1/\b^2$. In a 
relativistic context, this would give a realization of the DSR axioms [4]. Actually, 
several DSR models imply an upper bound for the spectrum of the momentum.

Let us again consider the one-dimensional model. 
The \ur are
$$\D x\D p\ge\ha\left|\,\langle[\hat x,\hat p]\rangle\right|={\hbar\over2}\,
\left|1-\b^2(\D p)^2-\b^2\brah p\ket^2\right|.\eqno(3.21)$$
In this case there is no minimal position uncertainty. However, contrary to standard quantum
mechanics, the states with vanishing position uncertainty have finite momentum uncertainty, 
given by $\D p=1/\b$. 

In analogy with the previous case, the position and momentum operators $\hat p$ and $\hat x$
can be realized on functions defined on a momentum space parametrized by $P$, as
$$\hat p\,\q(P)=p\,\q(P)={P\over\pref}\ \q(P),\qquad\hat x\,\q(P)=i\hbar\pref\ 
{\de\q(P)\over\de P}.\eqno(3.22)$$
The scalar product for which the operators (3.22) are symmetric is given by
$$(\q,\f)=\int_{-\inf}^{\inf}{dP\over\pref}\ \q^*(P)\,\f\,(P).\eqno(3.23)$$

The position eigenfunctions $\q_x$ with eigenvalue $x$ are determined by the equation
$$i\hbar\pref\ {\de \q_x\over\de P}=x\,\q_x,\eqno(3.24)$$
whose solution is 
$$\q_x=C\ {\rm exp}\left[-{ix\over\hbar\b}\,\arcsh\b P\right].\eqno(3.25)$$
The properties of these eigenfunctions are analogous to those holding in standard
quantum mechanics. They are not normalizable, since
$$|\q_x|^2=|C|^2\int_{-\inf}^{\inf}{dP\over\pref}\to\inf.\eqno(3.26)$$
Nevertheless, setting $C=1/\sqrt{2\p\hbar}$, the scalar product of position eigenstates 
gives
$$(\q_x,\q_{x'})=\d(x-x').\eqno(3.27)$$

The functions (3.25) are now physical eigenstates, with vanishing $\D x$, and finite \ev for 
$p$ and $p^2$, 
but, contrary to ordinary quantum mechanics, they exhibit finite momentum uncertainty, 
$\D p=1/\b$, as can be readily checked.
\bigbreak

\section{3.3 The pro-Snyder model}
As we have seen for the classical theory, the \cor of the anti-Snyder model can be realized 
also in an alternative way, by defining
$$\hat x_i=\sqrt{\b^2\hat P_k^2-1}\ \hat X_i,\qquad 
\hat p_i={\hat P_i\over\sqrt{\b^2\hat P_k^2-1}}\eqno(3.28)$$
where $\hat X_i$ and $\hat P_i$ obey canonical \cor and the spectrum of $P_i$ satisfies 
the bound $P_k^2>1/\b^2$. Also the momentum $p_i$ has a lower bound, $p_k^2>1/\b^2$. 
This setting is similar to the previous one, and its relativistic generalization
may be considered as a different kind of DSR, with lower bounds for the momenta.

Since the \cor are the same as in the anti-Snyder case, also the \ur enjoy the same 
properties.
In one dimension, the position and momentum operators $\hat p$ and $\hat x$
can be realized on functions defined on a momentum space parametrized by $P$, as
$$\hat p\,\q(P)=p\,\q(P)={P\over\pren}\ \q(P),\qquad\hat x\,\q(P)=i\hbar\pren\ 
{\de\q(P)\over\de P},\eqno(3.29)$$
and a suitable scalar product is given by
$$(\q,\f)=\int_{|P|\ge1/\b}\ {dP\over\pren}\ \q^*(P)\,\f\,(P),\eqno(3.30)$$

The position eigenfunctions $\q_x$ with eigenvalue $x$ are determined by the equation
$$i\hbar\pren\ {\de \q_x\over\de P}=x\,\q_x$$
whose solution is 
$$\q_x=C\ {\rm exp}\left[-{ix\over\hbar\b}\,\arcch\b P\right].\eqno(3.31)$$
The eigenfunctions are not normalizable, since
$$|\q_x|^2=|C|^2\int_{|P|\ge1/\b}{dP\over\pren}\to\inf,\eqno(3.32)$$
but setting $C=1/\sqrt{2\p\hbar}$, the scalar product of position eigenstates gives 
$$(\q_x,\q_{x'})=\d(x-x').\eqno(3.33)$$

However, contrary to the previous case, the expectation values of $\hat p$ and $\hat p^2$ 
diverge for the states (3.31), due to the singularity at $P^2=1/\b^2$. This model
appears therefore to be unviable.

\section{3.4 Quantum symmetries}
The invariance of the classical model under rotations and translations can be extended to 
the quantum case.

The rotations are generated by 
$$\hat J_{ij}=\hat x_i\hat p_j-\hat x_j\hat p_i=i\hbar\left(P_j{\de\over\de P_i}-
P_i{\de\over\de P_j}\right).\eqno(3.34)$$
and act in the standard way. In particular, the spectrum of $\hat J_{ij}$ is the same as in 
ordinary quantum mechanics.

As in the classical Snyder model, also in the quantum case the definition of translations is 
ambiguous, since their action is nonlinear. In particular, one may take as generators of 
translations the momentum operators $\hat p_i$, whose action on the positions operators is 
deformed, due to the \cor (3.1).
Analogously, in some problems it may be more convenient to identify the translation 
generators with the $\hat P_i$. 

Anyway, the \cor (3.1) transform covariantly under the symmetries defined above.

\section{3.5 Position eigenstates in three dimensions}
The analysis of the position operators of the previous sections easily generalizes to three 
dimensions. It must be taken into account, however, that the position operators $\hat x_i$ do 
not commute. For a free particle, the most relevant observable is the rotation-invariant 
radial coordinate $\hat r=\sqrt{\hat x_i^2}$, and therefore we adopt radial coordinates. 

As before, we work in a momentum representation. 
A basis of operators for one-particle states is given by the radial momentum 
$\hat p_r=\sqrt{\hat p_i^2}$ and the angular momentum $\hat L_i=\e_{ijk}\hat J_{jk}$.

Let us consider for example the Snyder model. Defining 
$$\hat p_r\q={P_r\over\prep}\ \q,\qquad
\hat r\,\q=i\hbar\prep\left({\de\over\de P_r}+{1\over P_r}\right)\q,\eqno(3.35)$$
one has
$$[\hat r,\hat p_r]=i\hbar(1+\b^2p_r^2).\eqno(3.36)$$ 
The uncertainty relations for the radial coordinates are therefore identical to those of the 
one-dimensional particle and enjoy the same properties.

Since the angular momentum action is the standard one,
the wave function for a free particle can be expanded in spherical harmonics,
$$\q_{rlm}(P_r,P_\h,P_\f)=\q_r(P_r)\, Y_{lm}(P_\h,P_\f),\eqno(3.37)$$
and we only need to investigate the radial functions. Their scalar product can be 
defined as
$$(\q_r,\f_r)=\int_0^{1/\b}{P_r^2\,dP_r\over\prep}\ \q_r^*(P_r)\,\f_r\,(P_r).\eqno(3.38)$$

Since $\hat p$ acts by multiplication, its spectrum is trivial.
The spectrum of the radial position operator is instead obtained from the equation
$$i\hbar\prep\left({\de\over\de P_r}+{1\over P_r}\right)\q=r\q,\eqno(3.39)$$
whose solution is 
$$\q_x(P_r)=C\ {\ex^{-{ir\over\hbar\b}\,\arcsin\b P_r}\over P_r},\eqno(3.40)$$
with $C$ an integration constant.
The $1/P_r$ factor in the eigenfunctions cancels the $P_r^2$ in the scalar product (3.38), 
so that the orthogonality properties of the eigenfunctions are the same as in one dimension, 
except that the integrals are restricted to positive values of $P_r$. We shall therefore 
not repeat the discussion of sect.\ 3.1. We just recall that these are not physical states,
since the \ev of the energy diverges.

One can however introduce maximally localized states, that satisfy the equation
$$\left[i\hbar\prep\left({\de\over\de P_r}+{1\over P_r}\right)-r+i\hbar\b^2{P_r\over\prep}
\right]\f_r=0,\eqno(3.41)$$
which has solutions
$$\f_{x}=N\ {\prep\over P_r}\ \ex^{-{ir\over\hbar\b}\,\arcsin\b P_r}.\eqno(3.42)$$
Also the properties of the maximally localized eigenfunctions reproduce those of their 
one-dimensional analogues.

The basis adopted in this section also permits to immediately find the states that minimize 
the \ur between the position coordinates along different directions. In fact, from (3.1) it
follows that 
$$\D x_i\D x_j\ge{\hbar\b^2\over2}\ |\bra J_{ij}\ket|,\eqno(3.43)$$
and the states that minimize these \ur are those with vanishing angular momentum.

The previous discussion can be easily extended to the anti-Snyder model.

\section{3.6 The Schr\"odinger equation}
The \schr equation for free particles in momentum space is algebraic and in our case
gives rise to the dispersion relation, that relates the energy to the momentum. It reads
$${\hat p_k^2\over2m}\ \q={p_k^2\over2m}\ \q=E\q,\eqno(3.44)$$
yielding the usual relation between momentum and energy.
If one adopts instead the definition $H=\hat P_k^2/2m$, the equation becomes
$${\hat P_k^2\over2m}\ \q={1\over2m}\ {p_k^2\over1+\l p_k^2}\ \q=E\q,\eqno(3.45)$$
leading to a deformed dispersion relation. In the following,
we shall not consider this possibility.
\medskip
A less trivial problem is given by the one-dimensional harmonic oscillator, with 
Hamiltonian
$$H={\hat p^2\over2m}+{m\o_0^2\hat x^2\over2}.\eqno(3.46)$$
For unit mass, the \schr equation for the Snyder oscillator is, in the representation (3.7),
$${d^2\q\over dP^2}-{\b^2P\over1-\b^2P^2}\ {d\q\over dP}-{1\over\hbar^2\o_0^2}
\left[{P^2\over(1-\b^2P^2)^2}-{2E\over1-\b^2P^2}\right]\q=0,\eqno(3.47)$$
with $P^2<1/\b^2$. In terms of a variable $\bar P=\arcsin\b P$, this becomes the
standard \schr equation for a potential 
$$V={1\over\o_0^2}\ \tan^2\bar P,\eqno(3.48)$$
which coincides with classical potential of sect.\ 2.3.

In order to find the explicit solution of eq.\ (3.47), it is however more convenient to 
define the variable $z=(1+\b P)/2$, in terms of which the equation can be written in the 
standard hypergeometric form
$${d^2\q\over dz^2}+{z-\ha\over z(z-1)}\ {d\q\over dz}-
\left[{\m(z-\ha)^2\over z^2(z-1)^2}+{\e\over z(z-1)}\right]\q=0,\eqno(3.49)$$
with $\m=1/\hbar^2\o_0^2\b^4$, $\e=2E/\hbar^2\o_0^2\b^2$. The solution reads
$$\q=\const\times(1-\b^2P^2)^{(1+\sqrt{1+4\m})/4}\ F\left(a,b,c;{1+\b P\over2}\right),
\eqno(3.50)$$
with 
$$a=\ha(1+\sqrt{1+4\m})-\sqrt{\m+\e},\quad b=\ha(1+\sqrt{1+4\m})+\sqrt{\m+\e},\quad
c=1+\ha\sqrt{1+4\m}\,.$$

We require that $\q$ vanish at $P=\pm1/\b$, i.e.\ at $z=0,1$. This occurs when $a=-n$ or 
$b=-n$. In both cases, 
$$\e=(n+\ha)(1+\sqrt{1+4\m})+n^2,\eqno(3.51)$$ 
namely,
$$\eqalignno{E&=\hbar\o_0\left[\left(n+\ha\right)\left(\sqrt{1+{\hbar^2\o_0^2\b^4\over4}}+
{\hbar\o_0\b^2\over2}\right)+{\hbar\o_0\b^2\over2}\,n^2\right]&\cr
&\approx\hbar\o_0\left[n+\ha+{\hbar\o_0\b^2\over2}\left(n^2+n+\ha\right)\right].
&(3.52)}$$
Hence corrections of order $\hbar\o_0\b^2$ occur in the spectrum of the harmonic oscillator,
and the relation between the \gs energy and $\o_0$ is no longer linear.
The expression (3.52) for the energy is similar to the one obtained in [12] for a different 
\rep of the commutation relations, but there are some differences in the numerical 
coefficients. 

\medskip
The same calculation can be performed for the anti-Snyder oscillator.
The \schr equation reads now
$${d^2\q\over dP^2}+{\b^2P\over1+\b^2P^2}\ {d\q\over dP}-{1\over\hbar^2\o_0^2}
\left[{P^2\over(1+\b^2P^2)^2}+{2E\over1+\b^2P^2}\right]\q=0.\eqno(3.53)$$
Again one may define a new variable $\bar P=\arcsh\b P$, in terms of which (3.53) becomes the
standard \schr equation for a potential identical to that obtained for the classical motion,
$$V={1\over\o_0^2}\ \tanh^2\bar P,\eqno(3.54)$$
and therefore bound states are possible for $0\le E<\b^2/2$ (see fig.\ 3). The latter 
inequality is also a consequence of the bound on the momentum $p^2<1/\b^2$.

Defining $z=(1+i\b P)/2$, eq.\ (3.53) can be written in the form of a hypergeometric 
differential equation, 
$${d^2\q\over dz^2}+{z-\ha\over z(z-1)}\ {d\q\over dz}-
\left[{\m(z-\ha)^2\over z^2(z-1)^2}-{\e\over z(z-1)}\right]\q=0,\eqno(3.55)$$
with $\m=1/\hbar^2\o_0^2\b^4$, $\e=2E/\hbar^2\o_0^2\b^2$. The solution is given by
$$\q=\const\times(1+\b^2P^2)^{(1-\sqrt{1+4\m})/4}\ F\left(a,b,c;{1+i\b P\over2}\right),
\eqno(3.56)$$
with 
$$a=\ha(1-\sqrt{1+4\m})-\sqrt{\m-\e},\quad b=\ha(1-\sqrt{1+4\m})+\sqrt{\m-\e},\quad
c=1-\ha\sqrt{1+4\m}\,.$$
The function $\q$ is regular at infinity if $a=-n$ or $b=-n$. In both cases,
$$\e=\left(n+\ha\right)\left(\sqrt{1+4\m}-1\right)-n^2,\eqno(3.57)$$ 
i.\ e.
$$\eqalignno{E&=\hbar\o_0\left[(n+\ha)\left(\sqrt{1+{\hbar^2\o_0^2\b^4\over4}}-{\hbar\o_0\b^2\over2}
\right)-{\hbar\o_0\b^2\over2}\,n^2\right]&\cr
&\approx\hbar\o_0\left[n+\ha-{\hbar\o_0\b^2\over2}\left(n^2+n+\ha\right)\right].&
(3.58)}$$
As it could have been guessed, the energy spectrum is simply the analytic continuation of the 
Snyder one for $\b\to i\b$, with analogous properties. However, an important difference arises. 
In the present case, the energy (3.58) becomes negative for large $n$. In order to preserve 
the bound $E\ge0$, one must impose that $n\le\sqrt{\m+{1\over4}}+\sqrt\m-\ha$, and hence only 
a finite number of energy levels are present.

\medskip
For the pro-Snyder model, the \schr equation takes the form 
$${d^2\q\over dP^2}+{\b^2P\over\b^2P^2-1}\ {d\q\over dP}-{1\over\hbar^2\o_0^2}
\left[{P^2\over(\b^2P^2-1)^2}-{2E\over\b^2P^2-1}\right]\q=0,\eqno(3.59)$$
with $P^2>1/\b^2$. In terms of the variable $\bar P=\arcch\b P$, this becomes the
standard \schr equation for a potential 
$$V={1\over\o_0^2}\ \coth^2\bar P,\eqno(3.60)$$
as in the classical case. It is evident that this potential does not admit bound states (see 
fig.\ 5). 
An explicit solution of (3.59) can still be obtained in terms of hypergeometric functions, but 
we shall not report it here.

\section{3.7 Quantization of area}
Another interesting implication of the Snyder model is that it gives rise to the quantization 
of areas, as was first shown in ref.\ [13].

This fact can be deduced by noting that every pair of spatial coordinates and their commutator 
satisfy an $so(3)$ algebra. For example,
$$[\hL_1,\hL_2]=i\hL_3,\qquad[\hL_3,\hL_1]=i\hL_2,\qquad[\hL_2,\hL_3]=i\hL_1,\eqno(3.61)$$
where $\hL_1=\hx_1/\b$, $\hL_2=\hx_2/\b$ and $\hL_3=\hat J_{12}$.

For a disc in the $x_1x_2$ plane, the area operator is defined as 
$$\hat A=\p(\hat x_1^2+\hat x_2^2)=\p\b^2(\hat L^2-\hat L_3^2),\eqno(3.62)$$
where  $\hL^2=\hL_1^2+\hL_2^2+\hL_3^2$ is the Casimir operator of $so(3)$. 
Defining as usual
$$\hL^2|\,lm\ket=l(l+1)|\,lm\ket,\qquad\hL_3|\,lm\ket=m|\,lm\ket,\eqno(3.63)$$
for $l=0,1,\dots$, $|m|\le l$, the spectrum of $\hat A$ follows immediately\footnote{$^2$}
{We do not consider spinorial representations.},
$$\hat A|\,lm\ket=\p\b^2[l(l+1)-m^2]|\,lm\ket,\eqno(3.64)$$
and is therefore discrete. 

In the anti-Snyder case, instead, the \cor become
$$[\hL_1,\hL_2]=-i\hL_3,\qquad[\hL_3,\hL_1]=i\hL_2,\qquad[\hL_2,\hL_3]=i\hL_1,\eqno(3.65)$$
and the operators span an $so(2,1)$ algebra, which is not compact and hence admits a 
continuous spectrum.

In fact, the area operator is now
$$\hat A=\p(\hat x_1^2+\hat x_2^2)=\p\b^2(\hat L^2+\hat L_3^2),\eqno(3.66)$$
where $\hL^2=\hL_1^2+\hL_2^2-\hL_3^2$ is the Casimir operator of $so(2,1)$. 
The algebra $so(2,1)$ admits both continuous and discrete spectra [17]. The continuous
spectrum is given by 
$$\hL^2|\,lm\ket=l^2|\,lm\ket,\qquad\hL_3|\,lm\ket=m|\,lm\ket,\eqno(3.67)$$
with $l>0$, $m=0,\pm1,\pm2,\dots$, 
and hence
$$\hat A|\,lm\ket=\p\b^2[l^2+m^2]|\,lm\ket.\eqno(3.68)$$

The discrete series is given instead by 
$$\hL^2|\,lm\ket=l(1-l)|\,lm\ket,\qquad\hL_3|\,lm\ket=m|\,lm\ket,\eqno(3.69)$$
for $l=1,2,\dots$, $|m|\ge l$.
In this case,
$$\hat A|\,lm\ket=\p\b^2[l(1-l)+m^2]|\,lm\ket.\eqno(3.70)$$

These results can be easily extended to the volume operator [13]. We conclude that, while
in the Snyder model positions, areas and volumes are quantized, this does not necessarily 
occur in the anti-Snyder case.

\section{4. CONCLUSIONS}
We have studied the classical and the quantum mechanics of free particles in the 
nonrelativistic version of the Snyder model. 
The calculations are based on the existence of a nonlinear 
transformation relating the Snyder phase space \coo to canonical coordinates.

As in the relativistic case, that will be treated in detail elsewhere,
the results strongly depend on the sign of the \cc in the 
defining \pb (or \cor in the quantum case).
In particular, it turns out that the spectrum of length, area and volume operators in the
quantum model is not necessarily discrete, in spite of the noncommutativity of the geometry.
Furthermore, the range of definition of the momenta depends on the sign of the coupling constant.

Also interesting is the behavior of the one-dimensional harmonic oscillator: both in the 
classical and in the quantum cases, its frequency is energy dependent. Moreover, if the \cc 
is negative, the quantum spectrum has a finite number of eigenvalues. 

Of course, the extension of the present investigations to the relativistic
theory is of special importance. This gives rise to interesting conceptual problems [14], in 
particular in the quantum mechanical case, where a field theory should be defined.
The dependence of the oscillator frequency on the energy may have important implications for 
the quantum fields.

The present analysis may also be extended to the case of Yang's model [18], 
where the background space is no longer flat, but has constant curvature, and in particular 
to its nonlinear realization, called triply special relativity [19,15]. In this case, however, 
the discovery of a transformation relating the noncanonical phase space variables to
canonical ones appears to be more problematic.

\beginref
\ref [1] H.S. Snyder, \PR{71}, 38 (1947).
\ref [2] M.R. Douglas and N.A. Nekrasov, \RMP{73}, 977 (2001);
R.J. Szabo, \PRep{378}, 207 (2003).
\ref [3] S. Doplicher, K. Fredenhagen and J.E. Roberts, \PL{B331}, 39 (1994).
For a review, see L.J. Garay, \IJMP{A10},145 (1995).
\ref [4] G. Amelino-Camelia, \PL{B510}, 255 (2001), \IJMP{D11}, 35 (2002).
For a recent review, see G. Amelino-Camelia, Symm.\ {\bf 2}, 230 (2010).
\ref [5] A. Granik, \hep{0207113}; 
S. Mignemi, \PR{D68}, 065029 (2003); 
S. Ghosh and P. Pal, \PR{D75}, 105021 (2007).
\ref [6] J. Kowalski-Glikman and S. Nowak, \IJMP{D13}, 299 (2003).
\ref [7] S. Mignemi, \PL{B672}, 186 (2009).
\ref [8] G. Jaroszkiewicz, \JoP{A28}, L343 (1995).
\ref [9] J.M. Romero and A. Zamora, \PR{D70}, 105006 (2004).
\ref [10] R. Banerjee, S. Kulkarni and S. Samanta, \JHEP{05}, 077 (2006).
\ref [11] M.V. Battisti and S. Meljanac, \PR{D79}, 067505 (2009).
\ref [12] A. Kempf, G. Mangano and R.B. Mann, \PR{D52}, 1108 (1995).
\ref [13] J.M. Romero and A. Zamora, \PL{B661}, 11 (2008).
\ref [14] E.R. Livine and D. Oriti, \JHEP{0406}, 050 (2004).
\ref [15] S. Mignemi, \AdP{522}, 924 (2010).
\ref [16] G. Veneziano, Europhys. Lett. {\bf2}, 199 (1986);
M. Maggiore, \PL{B304}, 63 (1993).
\ref [17] B.G. Wybourne, {\it Classical groups for physicists}, Wiley, New York 1974.
\ref [18] C.N. Yang, \PR{72}, 874 (1947).
\ref [19] J. Kowalski-Glikman and L. Smolin, \PR{D70}, 065020 (2004);
C. Chryssomakolos and E. Okon, \IJMP{D13}, 1817 (2004).

\endref
\end